\documentclass[a4paper, 10pt, twocolumn]{article}

\usepackage[utf8]{inputenc}         
\usepackage[english]{babel}          

\usepackage{graphicx}               
\usepackage{tabularx}               
\usepackage{multirow}               
\usepackage{hyperref}

\usepackage[small,bf]{caption2}     
\usepackage{parskip}
\usepackage{titlesec}
\usepackage{amsmath}                

\titleformat{\section}{\normalfont\large\bfseries}{\thesection}{}{}
\titleformat{\subsection}{\normalfont\large\bfseries}{\thesection}{}{}
\titleformat{\paragraph}{\normalfont\bfseries}{\theparagraph}{}{}
\titlespacing{\section}{0pt}{6pt}{-1pt}
\titlespacing{\subsection}{0pt}{3pt}{-1pt}
\titlespacing{\paragraph}{0pt}{3pt}{-1pt}

\newcolumntype{Y}{>{\centering\arraybackslash}X}    

\begin{document}

\date{}                                         

\title{\vspace{-8mm}\textbf{\large
Goniometers are a Powerful Acoustic Feature for Music Information Retrieval Tasks}}

\author{
Tim Ziemer$^1$\\
$^1$ \emph{\small University of Hamburg, 20354 Hamburg,
Germany, E-mail: tim.ziemer@uni-hamburg.de
}} \maketitle
\thispagestyle{empty}           
\section*{Introduction}
\label{sec:Einleitung}
Music Information Retrieval (MIR) task reach from blind source separation over automatic music transcription to music recommendation. Musicologists often use acoustic features and machine learning methods as explorative tool to analyze ethnographical recordings \cite{cultures,som}, popular music \cite{poma,hiphop} and large music collections \cite{poma,fire}. One of the most-used acoustic features for music analysis are Mel-Frequency Cepstral coefficients (MFCCs), even though they neither correlate with any aspect of music perception nor music production and mixing \cite{mel}. To be explanatory and intuitive, a feature should relate to the way music is produced, or the way music is perceived. The goniometer is a potential candidate because music producers and mixing engineers consult it to monitor and tune spatial aspects of their music mix. The present work is an explorative study to evaluate the intuitiveness and usefulness of the goniometer as a feature for music information retrieval of mixed stereo music.
\section*{The Goniometer}
\begin{figure}[hbt]
    \begin{center}
        \includegraphics[width=8.6cm]{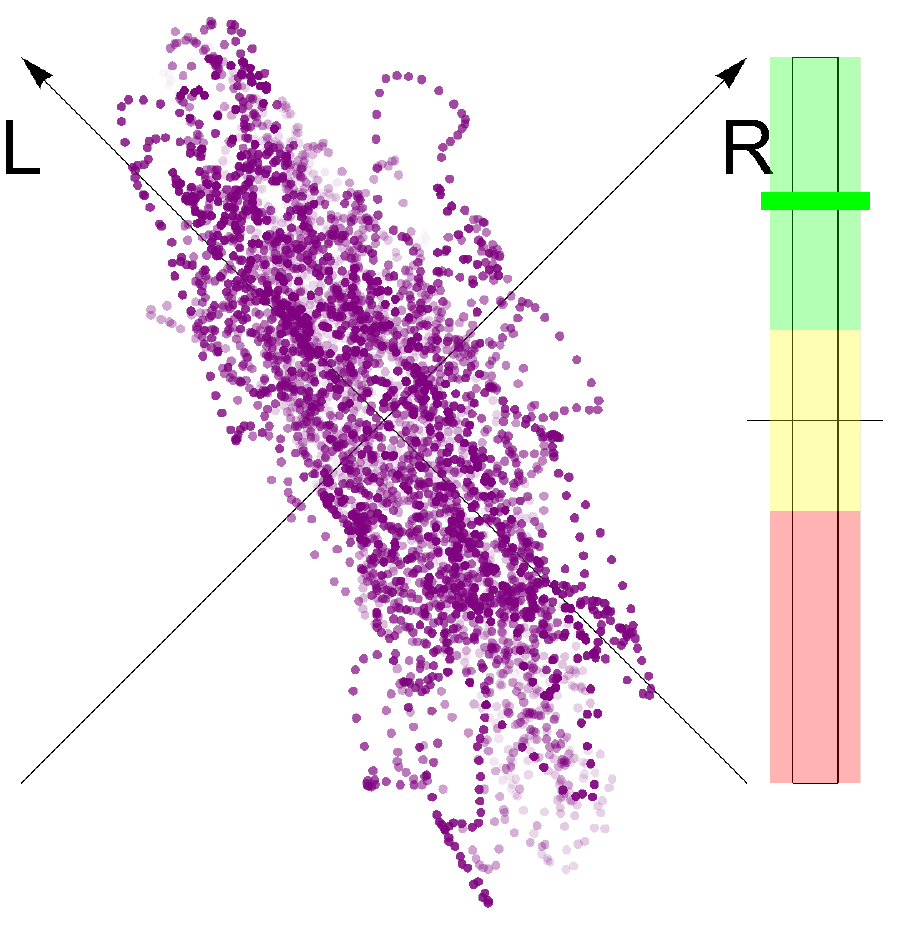}
    \end{center}
    \caption{Typical goniometer. The time signal of the left channel (L) is plotted over the time signal of the right channel (R) yielding a point cloud (purple). The normalized correlation of both signals is indicated by a bar on a bar diagram (right). Often, a high correlation is represented in green, while a high negative correlation is indicated in red.}
    \label{pic:gonio}
\end{figure}

Music producers and audio engineers know several recording and mixing techniques to control spatial aspects of a music production \cite{meister}. Many of these spatial aspects are monitored using a goniometer. A goniometer is illustrated in Fig. \ref{pic:gonio}. It consists of two meters, one point cloud that looks like Lissajous figures or a phase space diagram, and one bar diagram.

In the phase space diagram, the time series of the left channel is plotted over the time series of the right channel for all samples of a short-term buffer that may contain $4096$ samples. This buffer is also referred to as a time frame. Each sample pair is represented as a point, so all samples of a time frame produce a point cloud. Here, the time is represented as the opacity, i.e., the first samples from the buffer are transparent, while the last samples are opaque. The plot tends to be tilted by $-45^\circ$ for a simple reason: When a signal is routed to the first stereo channel, the signal will only be played by the left loudspeaker and the corresponding plot will look like a dotted line distributed along the L-axis. A mono signal equally routed to the left and the right channel produces a vertical dotted line. Due to the tilt, the plot gives a rough indication of signal panning in the stereo triangle. The Length of the line indicates the volume of the sound. All points in the upper square and in the lower square indicate that the signal is in phase, i.e., both loudspeakers are either positively or negatively elongated. Points in the left and the right square indicate that the signal is out of phase, i.e., that one loudspeaker is elongated positively while the other one is elongated negatively. A perfectly horizontal line through the axis origin is produced when the signal of the left and the right channel are perfectly out of phase. This exception is typically avoided, because it will lead to cancellation of bass frequencies in the typical listening setup, i.e., in a stereo triangle constellation. Furthermore, a mono mixdown of the signal will completely cancel the out-of-phase signal. Mono mixdowns are still quite common in at least two typical listening situations. Firstly, in nightclubs, when the loudspeakers are very far apart. Here, many mising engineers decide to mix both stereo channels at least partially. This way they avoid that a hard-panned instrument is only audible in a single corner of the dancefloor. Secondly, when music is played back over the internal loudspeakers of a mobile phone. In the good old days grandmother's portable radio was completely mono, while the younger generation sometimes listens to music through the mono tweeter of their mobile phone, or through portable bluetooth speakers whose stereo channels are installed in one tiny cabinet. The extreme opposite of a single line in the phase space is when the complete area is occupied by dots. This is only produced when each channel is fed with an individual white noise. Only when both white noises are normalized to the maximum sound pressure level, the complete space is filled with points. In all other cases, some of the regions in the field are unoccupied. Music producers and mixing engineers know several recording techniques and digital audio effects that widen the distribution of points to make instruments and ensembles sound wider and to make the listening experience more immersive, enveloping, spatial and vivid \cite{width}. At the same time, techniques exist to make wide distributions narrower, for example to increase the presence and the bass of single instruments, and to make them clearer and easier to localize. In summary, the width of the point cloud gives an indication of how spatial the overall mix is. The size of the point represents the volume. The overall direction of the distribution indicates the panning. In the plotted example, the sound from the left loudspeaker is louder, because a loud instrument panned to the left dominates the overall mix. Music producers and mixing engineers have their own philosophy concerning the appearance of the point cloud, e.g., how wide, how large, how fluctuating, and how tilted it should be, depending on the music material that they are mixing.

The second plot of the goniometer is a simple bar diagram. It indicates the normalized correlation of both buffered stereo channels. Consequently, the channel correlation can take values from $-1$ to $1$. Some goniometers also plot a short-term history, i.e., the highest and the lowest channel correlation during the last couple of seconds, or the minimum and maximum channel correlation that has been measured overall. Often, colors indicate whether a stereo mix is perfectly mono compatible (green), fairly mono compatible (yellow) or not mono compatible (red). Again, music producers and mixing engineers have their own philosophy concerning the region in which the bar should lie, the lowest position they will accept, and the dynamics of the bar position that they want, depending on the type of music that they mix.

\subsection*{The Goniometer Feature}
In practical use, music producers and mixing engineers consult the goniometer to observe the spatial aspects of the mix and to monitor how modifications through recording and mixing steps affect it. The goniometer is a music metering and monitoring tool. This procedure is rather qualitative than quantitative, i.e., practitioners evaluate the appearance of the two goniometer plots. To serve as an acoustic feature, the output of the goniometer has to be quantified.

To quantify the appearance of the point cloud, a simple box counting method is applied. The square that represents all possible coordinate pairs is divided into $20\times20$ sub-squares, as indicated in Fig. \ref{pic:gonio}. This yields $400$ sub-squares. Then, it is simply counted how many sub-squares are occupied by one or more points during one time frame. When we divide this number by $400$, we get a percentage. This percentage is stored for each analyzed time frame in a one-dimensional feature vector. In python, this can be computed efficiently by the following code for each time frame fs:
\begin{verbatim}
import numpy as np
bc = []#empty box counting feature vector
for i in range(fs):
   box[i]=[int(round(l[i]*10)),int(round(r[i]*10))]
bc.append(len(np.unique(box))/400)
\end{verbatim}
I call this feature \emph{Phase Scope} or \emph{Box Counting}. The variables $l$ and $r$ are arrays with a length of fs that contain  the time series of the left and the right channel.

To quantify the bar graph, its magnitude for each time frame is stored in a one-dimensional feature vector. In python this can easily be done, e.g., by:
\begin{verbatim}
import numpy as np
corr=[]#empty array for corelation feature vector
corr.append(np.min(np.corrcoef(l,r)))
\end{verbatim}
I call this feature \emph{Channel Correlation}.

The code for extracting the feature and writing feature vectors into a CSV-file is publically available on my GitHub repository \cite{github}.

In \cite{poma} we could already show that the two goniometer features can --- in combination with other recording studio features, like crest factor and root-mean-square energy --- predict which Disk Jockey would play which song. In this study, I evaluate the goniometer alone in an explorative way that reveals how the goniometer can serve as a feature for linear and nonlinear music classification and clustering tasks.

\section*{Method}
To evaluate the usefulness of the goniometer features, I chose an explorative approach. I analyzed some hundred songs. First, I imported stereo mp3 files and downsampled them to $22050$ Hz. Then, I took $500$ frames from the middle of the music piece. Each frame was $2048$ samples long and had no overlap, so I analyzed around $46$ seconds per file. Naturally, only stereo files were analyzed, as goniometers represent spatial aspects of stereophonic sound. To see how intuitive and explanatory the goniometer features were, I compared somewhat contrasting types of music. My first comparison was between $16$ songs from a meditation album by David Miles Huber, an American Composer and $15$ songs by Jimmy-J \& Cru-L-T, a UK duo that produced drum and bass music in the 1990s, also referred to as breakbeat hardcore. My second comparison was between $90$ classical music pieces and $93$ hip hop pieces, both in a broad sense. I visually inspected scatter plots of the two goniometer features for linear observations. When a linear separation between the two types of music was not possible, I trained a self-organizing map using the goniometer to see how explanatory the goniometer is when used with a nonlinear clustering method. The results were analyzed regarding causal relationships between the intention or sound ideal of the music producers/mixing engineers and the goniometer values.

\section*{Results \& Discussion}
\begin{figure}[hbt]
    \begin{center}
        \includegraphics[width=10cm]{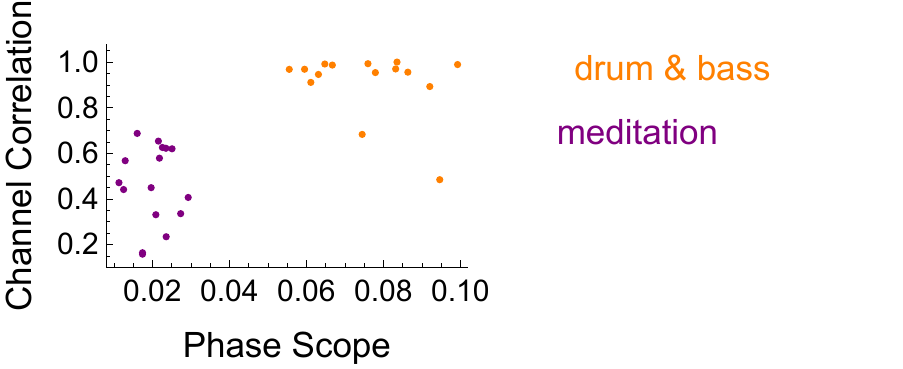}
    \end{center}
    \caption{Mean channel correlation over mean phase scope for drum and bass (orange) and mediation music (purple). These two goniometer features perfectly separate the two.}
    \label{pic:dblinear}
\end{figure}

The analysis results of the first comparison are plotted in Fig. \ref{pic:dblinear}. This scatter plot represents the feature space. Here, the drum \& bass and meditation music are very far apart and can be separated linearly. Every clustering algorithm, like k-means or self-organizing maps should be able to cluster them, and even linear classifiers, like single layer perceptrons, should be able to classify them with perfect accuracy. Note that in this case the phase scope is already sufficient to distinguish the two, while the channel correlation is not. This clear picture reveals a lot about the aesthetic ideal and the intention of the music producers. Jimmy-J \& Cru-L-T's drum and bass music is loud and compressed and has a high channel correlation to keep bass frequencies in phase for phat, percussive drums. David Miles Huber on the other hand produced much more dynamic music, with a wider, less correlated distribution. Of course, this is obvious to music experts. The fact that the goniometer features reflect this intention is evidence that the features may have a high explanatory value, as they help to find causal relationships between music and the producers' intention.

\begin{figure}[hbt]
    \begin{center}
        \includegraphics[width=9.6cm]{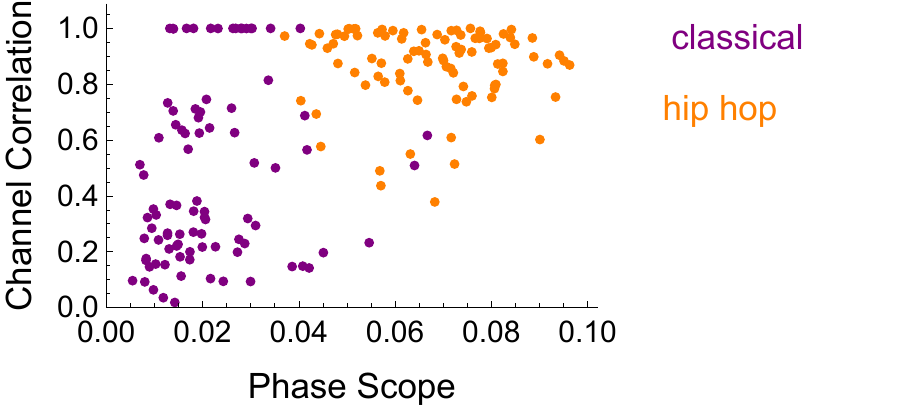}
    \end{center}
    \caption{Simply plotting the distribution of goniometer mean values shows obvious trends, classical music and hip-hop genres are not linearly separable.}
    \label{fig:dist}
\end{figure}

The scatter plot of the classical and hip hop music can be seen in Fig. \ref{fig:dist}. One can see similar trends as before, i.e., hip-hop music is found in the upper-right corner and classical music in the lower-left. But there are much larger overlaps along both dimensions. A linear separation between the two types of music is not possible with the goniometer features. 

\begin{figure}[hbt]
    \begin{center}
        \includegraphics[width=8.6cm]{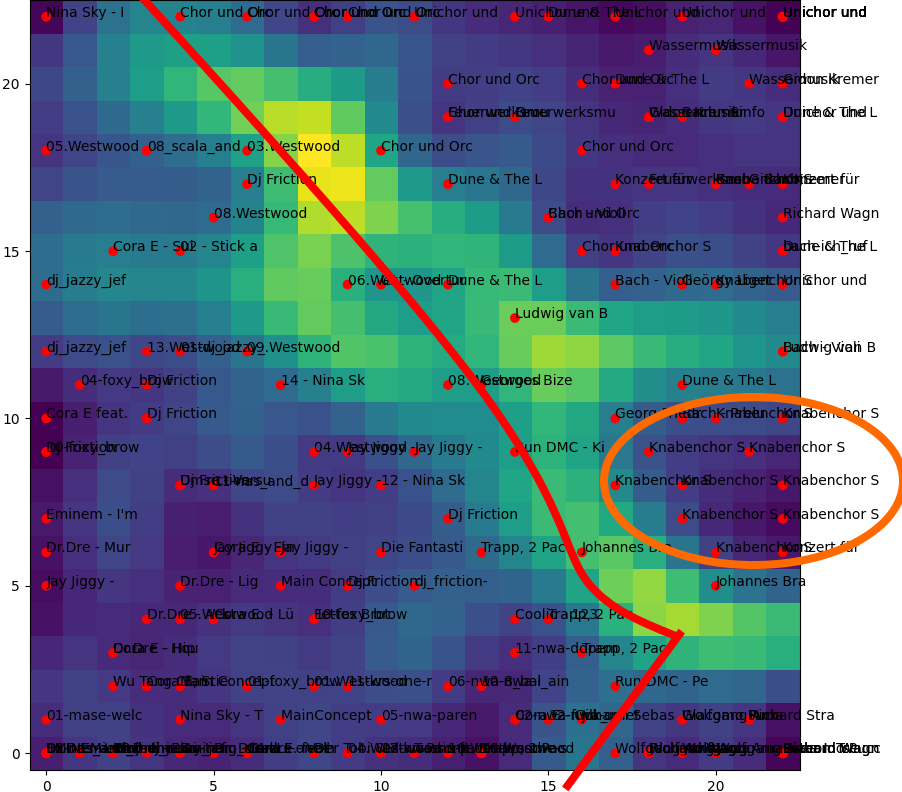}
    \end{center}
    \caption{Trained with the mean values of goniometer outputs, a self-organizing map perfectly separates hip-hop from classical music (red line) and even identifies a single album (orange oval).}
    \label{fig:som}
\end{figure}

To evaluate the usefulness of the goniometer features, they have been used to train a Self-Organizing Map (SOM) \cite{som}. SOMs are neural networks for unsupervised learning, i.e., clustering. A grid consisting of $23\times 23=529$ neurons was fitted to the training data in $500$ iteration steps, with a learning rate of $\eta=0.025$ and an initial neighborhood size of $\nu=25$, which is allowed to decrease linearly to $1$. The result is plotted in terms of a unified distance matrix in Fig. \ref{fig:som}. Here, dark blue fields represent clusters, yellow fields represent boarders. Each labeled red point represents one music piece. It can be seen that the SOM, trained on the goniometer features, clearly separates hip hop from classical music (red border) and even manages to cluster a single album by the Knabenchor St. Petersburg (orange oval mark) that is not visible in the linear feature space, Fig. \ref{fig:dist}. The SOM reveals that the goniometer features enable a nonlinear separation between the two analyzed types of music, and it even identified the Knabenchor St. Petersburg as a cluster, probably because the whole album was recorded in one session, with a fixed stereo microphone setup, yielding a homogeneous phase scope and channel correlation.

\begin{figure}[hbt]
    \begin{center}
        \includegraphics[width=8.6cm]{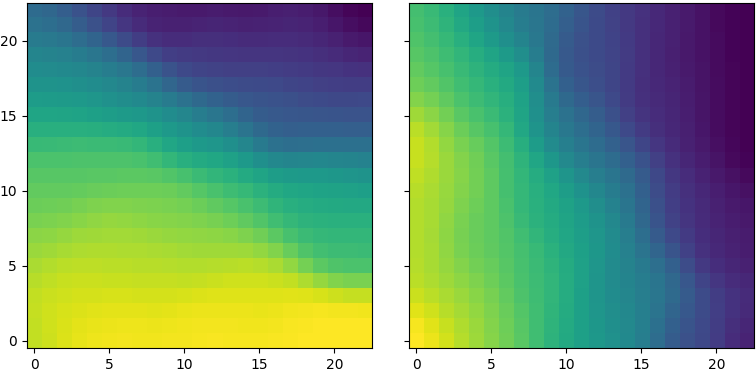}
    \end{center}
    \caption{Component matrices of the phase scope (left) and channel correlation (right) reveal that the two aspects of the goniometer are linearly independent, i.e., one is aligned vertically and the other one, horizontally.}
    \label{fig:comp}
\end{figure}

The two component matrices are illustrated in Fig. \ref{fig:comp}. They show how much each feature contributes to each field on the SOM. It can be seen that the weight of the phase scope is aligned from top to bottom, whereas the weight of the channel correlation is aligned from right-to-left. This orthogonality indicates that the two are orthogonal and therefore perfectly compliment each other. Orthogonal features are very important for machine learning, as redundancy reduces the accuracy of algorithms and redundancy reduction methods may be computationally costly.

\section*{Conclusion}
In this work, I suggest the goniometer as an explanatory acoustic feature for music information retrieval tasks. I define a way to quantify the qualitative view that music producers and mixing engineers have on the goniometer. Analyzing different types of music revealed that the goniometer yields intuitive, plausible results that serve to cluster different music styles explained by spatial and dynamic aspects of the music recording and production process. The phase scope and the channel correlation are orthogonal features that perfectly complement each other. In contrast to MFCCs, extracting the goniometer features needs very little computational power and the features are causally linked to the intention of the music producers. Additional analyses will reveal whether the goniometer features are also suitable to explore the other stereo sound files, such as podcasts, audio drama and the sound from movies, or recordings of the sound radiation characteristics of musical instruments, binaural recordings of architectural acoustics and soundscape recordings.

\end{document}